\documentclass[runningheads]{llncs}
\usepackage{graphicx}
\usepackage{subcaption}
\usepackage{xcolor}
\usepackage{indentfirst}
\usepackage{float}
\begin{document}
\title{Towards Automatic Scoring of Spinal X-ray for Ankylosing Spondylitis }
%
\titlerunning{Towards automatic scoring of spinal X-ray for ankylosing spondylitis}
%
\author{
Yuanhan Mo $^{\ast}$\inst{2}
Yao Chen $^{\ast}$ \inst{1} 
Aimee Readie\inst{1}
Gregory Ligozio\inst{1}\\
Thibaud Coroller $^{\dagger}$ \inst{1} 
Bart\l omiej W. Papie\.z $^{\dagger}$ \inst{2}
}
%
\authorrunning{Y. Mo et al.}
%
\institute{Novartis Pharmaceutical Company, East Hanover, NJ, USA \\
\email{\{yao.chen, aimee.readie, gregory.ligozio, thibaud.coroller\}@novartis.com} \and
Big Data Institute, University of Oxford, Oxford, UK\\
\email{yuanhan.mo@ndm.ox.ac.uk, bartlomiej.papiez@bdi.ox.ac.uk}\\
}

\maketitle              
\def\thefootnote{$\ast$}\footnotetext{Contribute equally}

\def\thefootnote{$\dagger$}\footnotetext{Contribute equally}

\begin{abstract}
Manually grading structural changes with the modified Stoke Ankylosing Spondylitis Spinal Score (mSASSS) on spinal X-ray imaging is costly and time-consuming due to bone shape complexity and image quality variations. 
In this study, we address this challenge by prototyping a 2-step auto-grading pipeline, called VertXGradeNet, to automatically predict mSASSS scores for the cervical and lumbar vertebral units (VUs) in X-ray spinal imaging.
The VertXGradeNet utilizes VUs generated by our previously developed VU extraction pipeline (VertXNet) as input and predicts mSASSS based on those VUs.
VertXGradeNet was evaluated on an in-house dataset of lateral cervical and lumbar X-ray images for axial spondylarthritis patients.
Our results show that VertXGradeNet can predict the mSASSS score for each VU when the data is limited in quantity and imbalanced.
Overall, it can achieve a balanced accuracy of 0.56 and 0.51 for 4 different mSASSS scores (i.e., a score of 0, 1, 2, 3) on two test datasets.
The accuracy of the presented method shows the potential to streamline the spinal radiograph readings and therefore reduce the cost of future clinical trials.

\end{abstract}
\section{Introduction}
X-ray imaging is one of the imaging modalities, utilized to monitor the structural progression of ankylosing spondylitis (AS), and this progression can be quantified by the modified Stoke Ankylosing Spondylitis Spinal Score (mSASSS) \cite{van2019modified}. 
Applying the mSASSS to each VU in an X-ray is usually performed manually by expert readers (radiologists) due to its complexity, making it a costly and tedious process. 
Moreover, such a manual process can introduce inter- and intra-reader variability in the final dataset. 
Therefore, to address these problems, we aim to propose an automatic pipeline for mSASSS grading.
\par
In this paper, we propose a 2-step auto-grading pipeline, VertXGradeNet, for estimating mSASSS scores from the given spinal X-ray images.
The proposed auto-grading pipeline was built on top of the previously developed VUs extraction pipeline \cite{vx_miua} and predicts mSASSS scores based on extracted VUs.
Then the proposed pipeline was validated utilizing clinical trial data from radiographic and non-radiographic axial spondyloarthritis patients.
\section{Method}
\begin{figure}[!t]
  \includegraphics[width=0.95\linewidth]{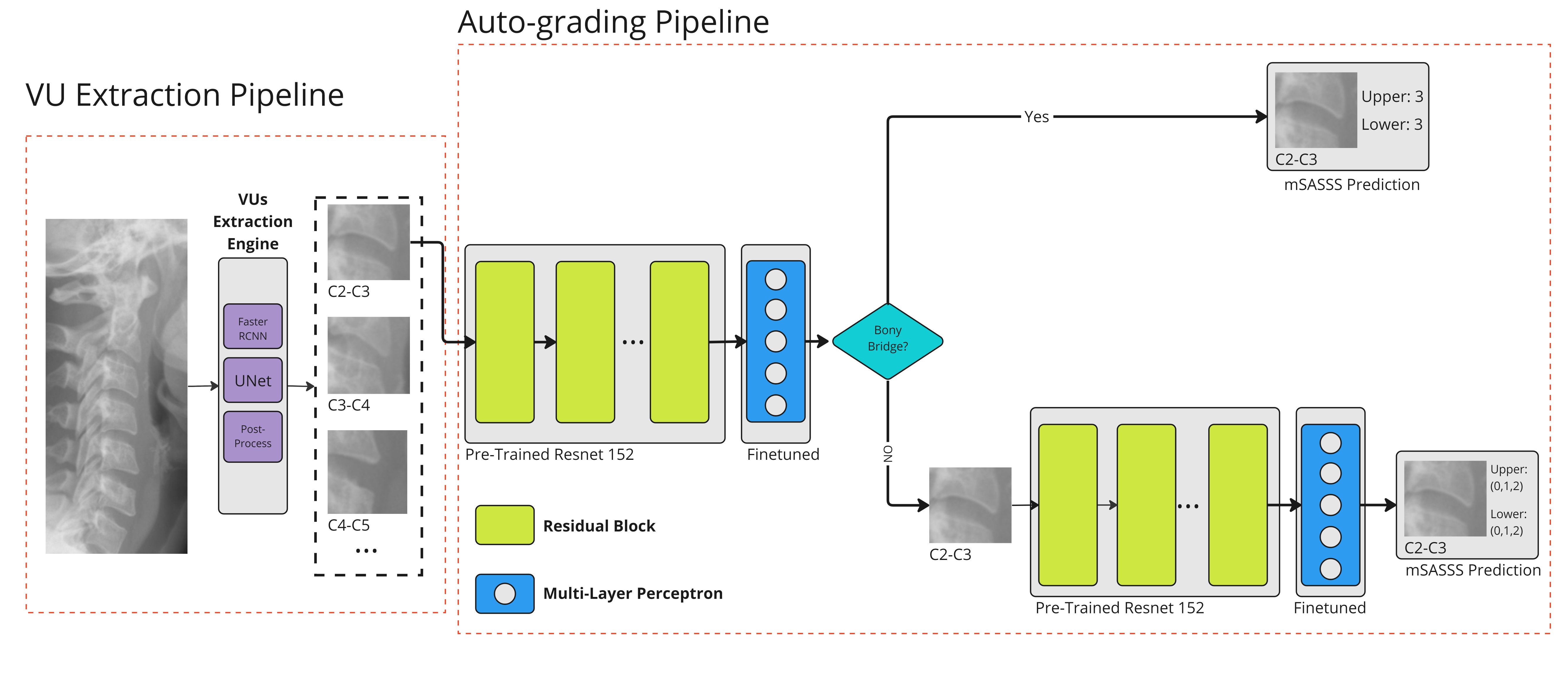}
  \caption{The initial step of vertebral units extraction Pipeline was developed previously \cite{vx_miua}. 
  The auto-grading pipeline was incorporated into the initial pipeline using the VUs as the input to predict the mSASSS.
  The auto-grading pipeline consists of two steps: the first step is to detect the bony bridge (mSASSS = 3) in the given VU, and the second step is to perform further mSASSS grading (i.e., mSASSS = 0,1 and 2) for a given VU without a bony bridge}
  \label{fig:pipeline}
\end{figure}
Inspired by work conducted by Koo et al. \cite{koo2022pilot}, a two-step grading pipeline is proposed (Figure~\ref{fig:pipeline}) which uses a deep neural network called ResNet~\cite{resnet} as the backbones. 
In the test phase, given a VU extracted by our in-house VU extraction pipeline, the first step of the proposed pipeline is to detect if a given VU has a bony bridge or not (i.e., mSASSS = 3 versus others).
If the given VU does not have a bony bridge, then the VU will proceed to the second step for further mSASSS prediction (i.e., mSASSS = 0,1 and 2).
In the second step, the pipeline will produce two predictions on both the upper and lower corners of the given VU (e.g. mSASSS = 0, 1, or 2) that is not classified as a bony bridge.
Finally, each given VU will have the estimated mSASSS given by the auto-grading pipeline.
\par
The ResNet152~\cite{resnet} is used twice in the two-step auto-grading pipeline.
In the first step, the ResNet 152 takes a VU as input and performs a binary classification, namely if the VU does or does not have a bony bridge.
In the second step, Another ResNet 152 takes the remaining VUs (mSASSS $\leq$ 2) as the inputs and predicts the rest of the mSASSS scores for both the upper and lower corners of a VU.
\par
In the training stage, a pre-trained ResNet 152 (on ImageNet) is fine-tuned on the extracted VUs/mSASSS scores from the MEASURE1 dataset.
The training process of the two stages of the auto-grading pipeline is not end-to-end.
Therefore, each step is trained independently.

\section{Experiments and Results}
{\bf Data.} The method was developed based on the anonymized datasets of secukinumab radiographic and non-radiographic axial spondyloarthritis clinical studies (MEASURE 1 \cite{baeten2015secukinumab} and PREVENT \cite{deodhar2021improvement}).
A total of 7239 and 9313 VUs were successfully extracted from the two studies. 

{\bf Experiments.} Extracted VUs were randomly split into 5 folds at the patient level for MEASURE 1 and performed 5-fold cross-validation.
The performance of the model, trained on MEASURE 1 data only, was evaluated on the PREVENT dataset.
The ground truth mSASSS scores for the training VUs were provided by the clinical trial. 
The detailed results are shown in Table \ref{prf1} and \ref{prf2}.



\begin{table}[h]
\centering
\begin{tabular}{l p{2cm} p{2cm} p{2cm} p{2cm}}
\hline
  mSASSS &  Precision & Recall & AUC(ROC) & F1-score\\ 
\hline
             0  & 0.934(0.010) & 0.918(0.007) & 0.897(0.011) & 0.926(0.010) \\
             1  & 0.200(0.097) & 0.240(0.103) & 0.809(0.076) & 0.218(0.096) \\
             2  & 0.390(0.069) & 0.300(0.020) & 0.849(0.021) & 0.332(0.026) \\
             3  & 0.654(0.067) & 0.800(0.023) & 0.959(0.009) & 0.718(0.034) \\
\hline
 Micro average  & 0.544(0.033) & 0.564(0.032) & 0.898(0.011) & 0.548(0.032) \\
 Macro average  & 0.860(0.014) & 0.856(0.014) & 0.879(0.020) & 0.856(0.014) \\
 \hline
\end{tabular}
\caption{Results of 5-fold cross-validation for MEASURE 1 dataset.}
\label{prf1}
\end{table}
\vspace{-0.8cm} 
\begin{table}[h]
\centering
\begin{tabular}{l r r r r r}
\hline
  mSASSS & Precision & Recall  & F1-score & AUC(ROC) & Support \\ 
 \hline
            0  & 0.99 & 0.95 & 0.97 & 0.825 & 15201 \\
            1  & 0.01 & 0.12 & 0.02 & 0.759 &    25 \\
            2  & 0.15 & 0.23 & 0.18 & 0.857 &   244 \\
            3  & 0.14 & 0.73 & 0.23 & 0.958 &    64 \\\hline
Micro average  & 0.32 & 0.51 & 0.35 & 0.826 & 15534 \\
Macro average  & 0.97 & 0.93 & 0.95 & 0.850 & 15534 \\
 
 \hline
\end{tabular}
\caption{Results for PREVENT dataset}
\label{prf2}
\end{table}

{\bf Conclusion.} We have prototyped a 2-step auto-grading pipeline for automatic mSASSS scoring. 
The current approach, which now can be considered as a benchmark, improves the grading performance compared to the preliminary results.
However, limited training samples and class imbalance issues still limit the current performance of the auto-grading pipeline. 
Thus, further analysis is required to address the aforementioned problems.

%
%
\newpage
\bibliographystyle{splncs04}
\bibliography{bibformiua}

\begin{thebibliography}{1}
\providecommand{\url}[1]{\texttt{#1}}
\providecommand{\urlprefix}{URL }
\providecommand{\doi}[1]{https://doi.org/#1}

\bibitem{baeten2015secukinumab}
Baeten, D., et~al.: Secukinumab, an interleukin-17a inhibitor, in ankylosing
  spondylitis. New England journal of medicine  \textbf{373}(26),  2534--2548
  (2015)

\bibitem{vx_miua}
Chen, Y., Mo, Y., Readie, A., Ligozio, G., Coroller, T., Papiez, B.W.:
  Vertxnet: Automatic segmentation and identification of lumbar and cervical
  vertebrae from spinal x-ray images (2022). \doi{10.48550/ARXIV.2207.05476},
  \url{https://arxiv.org/abs/2207.05476}

\bibitem{deodhar2021improvement}
Deodhar, A., Blanco, R., Dokoupilov{\'a}, E., Hall, S., Kameda, H., Kivitz,
  A.J., Poddubnyy, D., van~de Sande, M., Wiksten, A.S., Porter, B.O., et~al.:
  Improvement of signs and symptoms of nonradiographic axial spondyloarthritis
  in patients treated with secukinumab: primary results of a randomized,
  placebo-controlled phase iii study. Arthritis \& Rheumatology
  \textbf{73}(1),  110--120 (2021)

\bibitem{resnet}
He, K., Zhang, X., Ren, S., Sun, J.: Deep residual learning for image
  recognition. CoRR  \textbf{abs/1512.03385} (2015),
  \url{http://arxiv.org/abs/1512.03385}

\bibitem{koo2022pilot}
Koo, B.S., Lee, J.J., Jung, J.W., Kang, C.H., Joo, K.B., Kim, T.H., Lee, S.: A
  pilot study on deep learning-based grading of corners of vertebral bodies for
  assessment of radiographic progression in patients with ankylosing
  spondylitis. Therapeutic Advances in Musculoskeletal Disease  \textbf{14},
  1759720X221114097 (2022)

\bibitem{van2019modified}
Van Der~Heijde, D., Braun, J., Deodhar, A., Baraliakos, X., Landew{\'e}, R.,
  Richards, H.B., Porter, B., Readie, A.: Modified stoke ankylosing spondylitis
  spinal score as an outcome measure to assess the impact of treatment on
  structural progression in ankylosing spondylitis. Rheumatology
  \textbf{58}(3),  388--400 (2019)

\end{thebibliography}
%




\end{document}